\documentclass[twocolumn,showpacs,preprintnumbers,amsmath,amssymb,prl,floatfix]{revtex4}

\usepackage{psfig}
\usepackage{epsfig}
\expandafter\ifx\csname package@font\endcsname\relax\else
 \expandafter\expandafter
 \expandafter\usepackage
 \expandafter\expandafter
 \expandafter{\csname package@font\endcsname}%
\fi
\begin{document}


\title{Unconventional carrier-mediated ferromagnetism above room temperature in ion-implanted (Ga, Mn)P:C\\}

\author{N. Theodoropoulou}
\author{A. F. Hebard}%
 \email{afh@phys.ufl.edu}
\affiliation{%
Department of Physics, University of Florida, Gainesville, FL 32611-8440\\
}%
\author{M.E. Overberg}%
\author{C.R. Abernathy}%
\author{S.J. Pearton}%
\affiliation{
Department of Materials Science and Engineering, University of Florida, Gainesville, FL 32611-8440\\
}%

\author{S.N.G. Chu}
\affiliation{Bell Laboratories, Lucent Technologies, Murray Hill, NJ 07974\\
}%

\author{R.G. Wilson}
\affiliation{
Consultant, Stevenson Ranch, CA 95131\\
}%

\date{\today}

\begin{abstract}
Ion implantation of Mn ions into hole-doped GaP has been used to induce ferromagnetic
behavior above room temperature for optimized Mn concentrations near
3\,at.$\%$. The magnetism is suppressed when the Mn dose is increased or decreased
away from the 3\,at.$\%$ value, or when $n$-type GaP substrates are used. At low
temperatures the saturated moment is on the order of one Bohr magneton, and the
spin wave stiffness inferred from the Bloch-law $T^{3/2}$ dependence of the
magnetization provides an estimate $T_c$ = 385K of the Curie temperature that exceeds
the experimental value, $T_c$ = 270K. The presence of ferromagnetic clusters and
hysteresis to temperatures of at least 330K is attributed to disorder and proximity
to a metal-insulating transition.

\end{abstract}

\maketitle

The rather short interval between discovery of ferromagnetism in the Mn
doped III-V semiconductor GaAs \cite{Ohn01179} and the demonstration of unique phenomena
such as field-effect control of ferromagnetism \cite{Ohn00944}, efficient spin injection to produce
circularly polarized light \cite{Fie99787,Ohn99790}, and spin-dependent
resonant tunneling \cite{Ohn98951}, opens a rich
and varied landscape for technological innovation in magnetoelectronics \cite{Wol011488}.
Although the Curie temperature, $T_c = 110$K, of (Ga,Mn)As is spectacularly high, it is
not high enough, and any real breakthrough with respect to applications will
require diluted magnetic semiconductors (DMS) which exhibit ferromagnetism
above room temperature \cite{Wol011488}. Already there are reports of room temperature
ferromagnetism in the DMS chalcopyrite Cd$_{1-x}$Mn$_x$GeP$_2$ \cite{Med00949}, (Ga,Mn)N
formed by diffusion \cite{Ree013473}, and Co-doped ZnO \cite{Ued01988}
and TiO$_2$ anatase \cite{Cha013467,Mat01854}.

In this letter we report on carrier-mediated ferromagnetism in (Ga,Mn)P,
doped $p$-type with carbon (C), that persists to above room temperature. High $T_c$
behavior is only observed for samples that are prepared at an optimal Mn concentration
near 3\,at.$\%$ on heavily p-doped GaP substrates. When electron-doped substrates
are simultaneously implanted with the same Mn concentration, $T_c$ is
strongly suppressed. Hysteresis in the 3\,at.$\%$ material extends to temperatures well
above the temperature where the dc susceptibility, $\chi (T)$, diverges and the
``saturated'' moment is observed to have strong temperature dependence. These observations
suggest a percolation type of picture in which isolated ferromagnetic
magnetic clusters, immersed in a background of paramagnetic moments, grow in
size as the temperature is lowered until, at $T = T_c$, long-range order extends
through the whole system. The characteristics, which we have observed and will
describe below, thus reveal phenomenology that can be used to assess the relevance
of distinct yet complementary theoretical
viewpoints \cite{Die001019,Kon005628,Ber01107203,Lit015593}.

The use of GaP as a host semiconductor for magnetic dopants is advantageous:
it has a large bandgap (2.2eV), it is an important component of the AlGaInP
materials system, which is used for visible light-emitting diodes and high
speed electronics, and it is also nearly lattice-matched to Si, offering the
possibility of combining DMS with conventional Si circuitry. Epitaxial films of
(Ga,Mn)P have shown single-phase material up to Mn atomic concentrations of
9\% and hysteretic magnetic behavior that persists up to 250K \cite{Ove013128}.

Two types of GaP substrates were used in these experiments. The first
were bulk (100) substrates with nominally undoped $n$-type (10$^{16}$cm$^{-3}$) background
carrier densities and the second were 0.4$\mu$m-thick epi layers grown by Gas
Source Molecular Beam Epitaxy on top of GaP substrates. These layers were
heavily C-doped (p $\approx$ 10$^{20}$ cm$^{-3}$, 10$^{-2}$Ohm cm) using carbon
tetrabromide (CBr$_4$) as
the dopant source \cite{Abe95203}.
The samples were implanted with 250keV Mn$^+$ ions to doses
of 3-5$\times$10$^{16}$ cm$^{-2}$ at a temperature of ~350C to maximize
crystallinity \cite{Kuc0151}. These
conditions produce relatively flat Mn concentration profiles with peak volume
concentrations of 3 or 5\,at.\% over a projected $\approx$ 0.2$\mu$m depth in each sample.
Following a rapid thermal annael, the samples were characterized by transmission electron
microscopy (TEM), selected area diffraction pattern analysis (SADP),
and double crystal x-ray diffraction (DC-XRD) \cite{Ove01XXX}. No evidence of secondary
phase formation of GaMn or MnP was found. All magnetic measurements were
made with a Quantum Design SQUID magnetometer with the applied field parallel
to the film surface. Similar results were also obtained for perpendicular fields.

\begin{figure}[t]
\begin{center}
      \includegraphics[width=0.9\linewidth]{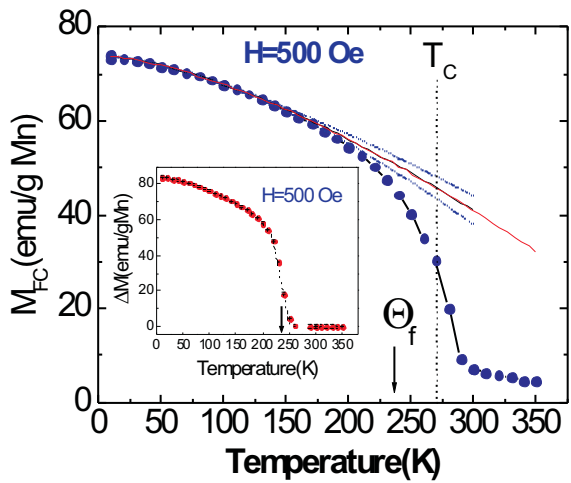}
\caption{\label{fig:fig1}
Temperature dependence of field-cooled (FC) magnetization (per gram
Mn) at at an applied field $H=500$Oe. 
At low temperatures a Bloch law ($T^{3/2}$) dependence (solid line) was
found. The dashed lines are 95\% confidence bands. The dashed line at $T_c = 270K$
marks the field independent inflection point (Curie temperature) and the vertical
arrows in the main panel and the inset mark the ferromagnetic Curie temperature,
$\Theta_f = 236K$. Inset: Temperature dependence of the difference $\Delta M$
between field cooled (FC) and zero-field cooled (ZFC) magnetizations taken in a
field of 500Oe.}
\end{center}
\end{figure}

\begin{figure}[bh]
\begin{center}
      \includegraphics[width=0.9\linewidth]{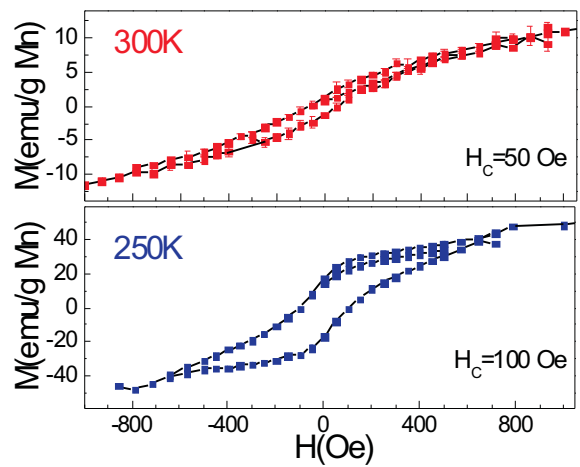}
\caption{\label{fig:fig2}
Magnetization loops showing moderate hysteresis with coercive field,
$H_c = 50$Oe (upper panel) at 300K and a stronger hysteresis with $H_c = 100$Oe
(lower panel) at 250K.
}
\end{center}
\end{figure}

Shown in Fig.~1 is a plot of the temperature-dependent magnetization for
the 3at.\% (Ga$_{0.94}$Mn$_{0.06}$P) sample cooled in a field of 500Oe. A temperature
independent diamagnetic contribution was subtracted from the background. In
contrast to (Ga,Mn)As \cite{Tan931565} the temperature-dependent magnetization
has the classic convex outwards shape but with a substantial tail extending to
higher temperatures.
The Curie temperature $T_c=270K$, indicated by the vertical dashed line,
is defined as the inflection point and does not shift in position for similar data sets
taken in different fields and sample orientations.

At low temperatures the saturated moment, $M_0 = g \mu_B S$, is calculated to
be about one Bohr magneton ($\mu_B$) per Mn ion, thus implying that for a $g$-factor of
2 the spin $S = 1/2$. This value is a factor of five less than $S = 5/2$ expected for the
half-filled $d$-band of divalent Mn, a discrepancy most likely arising because of
strong antiferromagnetic coupling among the more closely spaced randomly positioned Mn
ions \cite{Die001019}. The solid line is a three parameter fit to an expression of the
form, $M(T) = M_0 ( 1- \alpha T^{\eta})$, over the temperature range $T \leq 160$K. The observation
of a Bloch law dependence with exponent $\eta$ = 3/2 is expected for ferromagnets
and reflects the presence of long-wavelength thermally excited spin waves. The
dashed lines delineate the 95\% confidence band for the best-fit parameters
$\eta$ = 1.51(6) and $\alpha$ = 1.09(35)$\times$10$^{-4}$.

\begin{figure}[b]
\begin{center}
      \includegraphics[width=0.9\linewidth]{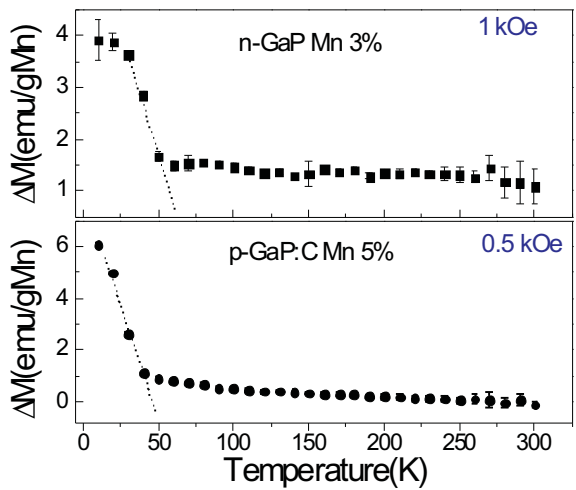}
\caption{\label{fig:fig3}
Temperature dependence of the difference between field-cooled (FC)
and zero-field cooled (ZFC) magnetizations (per gram Mn) for (upper panel)
$n$-type GaP implanted with 3at.\% Mn and (lower panel) $p$-type GaP:C implanted with
3at.\% Mn. The measuring fields are indicated in the upper right hand corners of
each panel.
}
\end{center}
\end{figure}

The inset of Fig.~1 shows the temperature-dependent difference
$\Delta M = M_{FC} - M_{ZFC}$ between field-cooled (FC) and zero-field-cooled (ZFC)
magnetization curves taken at 500Oe. This subtraction advantageously eliminates para- and
diamagnetic contributions and simultaneously indicates the presence of hysteresis
if the difference is non-zero. It should be noted that the knee of this curve occurs
at a lower temperature than the knee of the FC curve (main panel) and moves to
even lower temperature at higher measuring fields. This trend can be understood
by realizing that at a given temperature, $\Delta M = 0$ if the measuring field is higher
than the maximum field at which hysteresis is observed in $M(H)$ loops acquired at
the same temperature. Examples of such loops are shown in Fig.~2. Hysteresis
with a coercive field $H_c = 100$Oe is quite pronounced at 250K (lower panel) to
fields as high as 600Oe. At lower fields, hysteresis similar to that shown at 300K
in the upper panel of Fig.~2 extends to 330K.

The FC-ZFC subtraction is particularly effective when there are small
amounts of ferromagnetic material in the presence of a large diamagnetic and/or
paramagnetic background. This situation ensues when $n$-type rather than $p$-type
GaP is implanted with 3at.\% Mn (Fig.~3, upper panel) or $p$-type GaP is implanted
with 5at.\% Mn (Fig.~3, lower panel). These data show that the maximum FC-ZFC
difference, $\Delta M$, is reduced by a factor of 15-20 and the $T_c$
by a factor of $\approx$6, over the temperature range 300K to near 50K.
Accordingly, an optimized $T_c$ is extremely sensitive
to both the carrier and Mn concentrations. Furthermore, ferromagnetic impurity
phases such as MnGa ($T_c > 300K$) \cite{Tan931565} and MnP ($T_c = 291K$) \cite{Sha84361} are
not responsible,
since an increase of Mn leads to less, not more, magnetism, and the formation
and amount of such impurity phases should also be independent of hole-doping level.

Although our observations confirm conventional ferromagnetism at low
temperatures, the magnetic behavior is considerably more complicated at higher
temperatures as shown by the Fig.~4 plots of the temperature-dependent inverse
dc susceptibility $\chi^{-1}(T)$ (main panel) and saturation moment $M_s (T)$ (inset).
The data for these plots were obtained at selected temperatures by recording
$M_{ZFC} (H)$ to the
maximum available field of 5T. A temperature-dependent diamagnetic contribution
that varied less than 1\% over the temperature range studied was subtracted
from the data. We used a Langevin functional dependence to determine
$\chi = dM_{ZFC}/dH$ and $M_s (T)$ at each temperature. We identify the ferromagnetic
Curie temperature $\Theta_f = 236$K (vertical arrow) by a linear extrapolation
of $\chi^{-1} (T)$ to zero.
The hysteresis and accompanying saturated moment extend with diminishing magnitude
to substantially higher temperatures: 330K for the measured hysteresis and 390K
for a presumptuous linear extrapolation of $M_s (T)$ to zero. Using
the expression for the Curie constant, $C = g^2 \mu_{B}^{2} N_0 S(S+1)/3k_B$,
where $N_0$ is the
number of Mn atoms per gram Mn and $k_B$ is the Boltzmann constant, we calculate
from the slope of $\chi^{-1}(T)$, the result, $g^2 S(S+1) = 410$.
For $g = 2$ this implies that
$S = 10$, a spin larger by a factor of four than the $S = 5/2$ of isolated Mn$^{++}$ ions.
Such a large $S$ justifies the use of a Langevin (classical) rather than a Brillouin
(quantum) functional dependence.

\begin{figure}[b]
\begin{center}
      \includegraphics[width=0.9\linewidth]{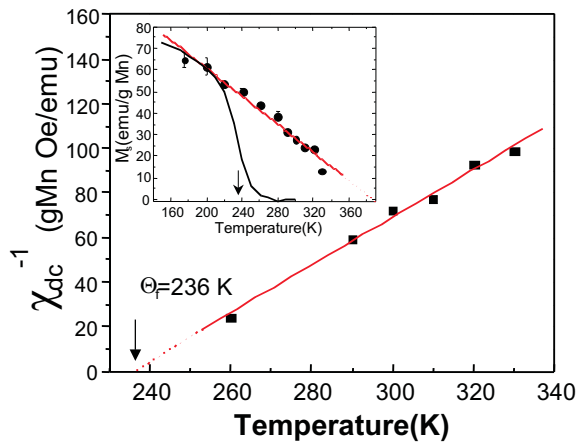}
\caption{\label{fig:fig4}
Temperature dependent inverse susceptibility extrapolated (dashed
line) to $\Theta_f = 236K$. The inset shows the temperature dependent
"saturated magnetization" derived from fits of the field-dependent magnetization at each
temperature to a Langevin function dependence as described in the text.
For comparison, the solid line reproduces $\Delta M(T)$ at 1000 Oe shown in the inset of Fig.~1.
}
\end{center}
\end{figure}

The data presented here not only show that room-temperature ferromagnetism is
obtainable in (Ga,Mn)P:C and but also provide a meaningful confrontation with
present theories
\cite{Die001019,Kon005628,Ber01107203,Lit015593} describing ferromagnetism in DMS.
The ferromagnetism is clearly carrier mediated
\cite{Die001019,Kon005628,Ber01107203,Lit015593} and is optimized for hole-doped substrates
with Mn concentrations near 3at.\%. Higher concentrations of Mn are deleterious
because of increased antiferromagnetic coupling \cite{Die001019} and a possible disorder-induced
self-compensation \cite{Lit015593}. At low Mn concentrations, RKKY interactions randomize
the sign of the exchange coupling between Mn spins, and spin-glass rather than
ferromagnetic behavior is expected \cite{Kon0111314}.

Our experimental determination of the prefactor $\alpha$ discussed above allows
a direct calculation of the spin wave stiffness $D$, a parameter which describes the
energy dispersion, $E = Dk^2$, of the spin wave excitations (magnons). We find from
the thermodynamic result \cite{Dys561230}, corrected by a spin-dependent prefactor \cite{Kon005628},
the value $D = 167meV \AA^2$. For comparison, $D = 281meV \AA^2$ for Fe. Substituting
$D = 167meV \AA^2$, $S = 5/2$, and $N_{Mn} = 1.5 \times 10^{21}$ cm$^{-3}$
for the volume density of Mn
ions into the expression \cite{Sch011550}, $k_B T_{c}^{coll} = (2S+1)D(6 \pi^2 N_{Mn} )^{2/3}$
gives an estimate $T_{c}^{coll} = 385K$ for the upper bound on $T_c$ due to long
wavelength collective fluctuations. This value is higher than the measured
$T_c = 270K$ and consistent with the
fact that mean field theory usually overestimates the ordering temperature.

At temperatures near and above $T_c$, clusters in a disordered medium seem
to be playing an important role \cite{Ber01107203,Lit015593}. The presence of spin clusters
is often associated \cite{Cul72} with a $\chi^{-1}(T)$ dependence, which approaches
zero with positive curvature
but which, at sufficiently high temperatures, has a linear dependence extrapolating
to the paramagnetic Curie temperature, $\Theta_p$.  It is therefore reasonable to argue that
at higher temperatures $\chi^{-1}(T)$ must eventually curve upwards and approach a slope
four times larger than shown in Fig.~4, corresponding to $S=5/2$ moments of the
Mn ions. If so, then a linear extrapolation to $\chi^{-1}(T=\Theta_p) = 0$ would give a
$\Theta_P$ considerably larger than $\Theta_f$. As $T$ increases beyond $T_c$ and the
ferromagnetic clusters decrease in size and number, there is a decreasing saturated
moment (Fig.~4, inset) since the remaining paramagnetic regions cannot be driven
into saturation without the assistance of internal Weiss fields. Disorder and proximity
to a metal-insulator transition, which may be responsible for the enhanced
$T_c$'s of the clusters \cite{Ber01107203}, may also give rise to a reduced spin stiffness
and the anomalously high Curie constant. Based on this evidence of ferromagnetic behavior
above room temperature, it is not unreasonable to expect that even higher $T_c$ can
be obtained by further optimization of the Mn and hole dopant concentrations.

The authors gratefully appreciate discussions with S. Arnason on experiment
and R. Bhatt, P. Kumar, and D. Maslov on theory. The work at UF is partially
supported by NSF DMR-0101438 (SJP) and DMR-0101856 (AFH), while
that of RGW is partially supported by ARO.
\bibliography{final}
\end{document}